# ANALYSIS OF BETA-DECAY RATES FOR Ag108, Ba133, Eu152, Eu154, Kr85, Ra226 AND Sr90, MEASURED AT THE PHYSIKALISCH-TECHNISCHE BUNDESANSTALT FROM 1990 to 1996

P.A. STURROCK[1*], E. FISCHBACH[2], AND J. JENKINS[3]

[1] Center for Space Science and Astrophysics, Stanford University, Stanford, CA 94305, USA
[2] Department of Physics and Astronomy, Purdue University, West Lafayette, IN 47907, USA
[3] Department of Nuclear Engineering, Texas A&M University, College Station, TX 77843, USA

* sturrock@stanford.edu

## ABSTRACT

We present the results of an analysis of measurements of the beta-decay rates of Ag108, Ba133, Eu152, Eu154, Kr85, Ra226, and Sr90 acquired at the Physikalisch-Technische Bundesanstalt from 1990 through 1995. Although the decay rates vary over a range of 165 to 1 and the measured detector current varies over a range of 19 to 1, the detrended and normalized current measurements exhibit a sinusoidal annual variation with amplitude in the small range 0.068% to 0.088% (mean 0.081%, standard deviation 0.0072%, an $11\sigma$ rejection of the zero-amplitude hypothesis) and phase-of-maximum in the small range 0.062 to 0.083 (January 23 to January 30). In comparing these results with those of other related experiments that yield different results, it may be significant that this experiment, at a standards laboratory, seems to be unique in using a $4\pi$ detector. These results are compatible with a solar influence, and appear not to be compatible with an experimental or environmental influence. It is possible that Ba133 measurements are subject also to a non-solar (possibly cosmic) influence.

## 1 . INTRODUCTION

It has long been known that beta-decay rates may be influenced by physical or chemical influences (Emery 1972; Hahn et al. 1976; Norman et al. 2001; Ohtsuki et al. 2004; Reifenschweiler 1998). In recent years, a number of investigators have presented evidence (summarized by Fischbach et al. 2009, and by Sturrock et al. 2014a) for the apparently spontaneous variability of beta-decay rates: Alburger et al. (1986); Baurov (2010); Baurov et al. (2007, 2010); Falkenberg (2001); Jenkins et al. (2009a); Lobashev et al. (1999); Namiot & Shnoll (2006); Parkhomov (2010a, 2010b, 2011, 2012); and Shnoll et al. (1998, 2000). Although there is as yet no accepted theory for explaining beta-decay variability, Fischbach et al. (2009) have drawn attention to the possibility proposed by Falkenberg (2001) that beta decays may be influenced by neutrinos. An alternative hypothesis, advanced by Shnoll et al. (1998, 2000) and by Namiot & Shnoll (2006), is that beta-decay fluctuations are due to some unknown cosmophysical process.

We here discuss the experimental results of Siegert, Schrader and Schoetzig of the Physikalisch-Technische Bundesanstalt (PTB) in Braunschweig, Germany, who have published the results of two extensive sets of measurements of beta-decay rates: the first (which we refer to as PTB-1) for Ag108, Ba133, Eu152, Eu154, Kr85, Ra226, and Sr90 acquired over the timeframe 1989.973 to 1995.993 (Siegert et al. 1998); and the second (which we refer to as PTB-2) for Ag108, Ba133, Cs137, Eu152, Eu154, Kr85, Ra226, and Sr90 acquired over the timeframe 1999.412 to 2008.871 (Schrader 2010). Dr H. Schrader at PTB has generously provided us with the results of all experimental measurements.

Both experiments used a pressurized 4π well-type ionization chamber, similar to those typically used in standards measurements. The experiments differed mainly as follows: for PTB-1, current measurements were made by means of a Townsend balance, whereas for PTB-2, current measurements were made by means of a Keithley electrometer. We have recently published the results of our analysis of the PTB-2 dataset (Sturrock et al. 2014b). We now offer the results of our analysis of the PTB-1 dataset.

Section 2 presents plots of the data before processing, and Section 3 presents plots of the data after processing. We discuss the results in Section 4.

## 2. DATA PLOTS BEFORE PROCESSING

We show, in Figures 1 through 7, plots of the (relative) current measurements for the seven nuclides. We see that they vary over a very wide range, from a low of 17 for Ag108 to a high of 326 for Eu152.

It is notable that some measurements show a more obvious annual oscillation than others. The oscillation is quite obvious for Ag108 and Ra226, but barely noticeable for the others. Ag108 and Ra226 are the nuclides with the longest half-lives (438 y and 1600 y, respectively). Table 1 shows the half-life estimates as derived from the PTB-1 data. (These estimates differ slightly from currently accepted values of half-lives.) A similar oscillation would not be as obvious in current measurements for nuclides with shorter half-lives.

Table 1. For each nuclide, the table lists the fitted half life, the minimum current, the maximum current, the mean current, and the ratio of the maximum and minimum currents.

| Nuclide | Half Life Fit (y) | Minimum Current | Maximum Current | Mean Current | Max/Min |
|---|---|---|---|---|---|
| Ag108 | 430 | 17.22 | 17.73 | 17.31 | 1.03 |
| Ba133 | 10.74 | 120.28 | 178.93 | 144.61 | 1.49 |
| Eu152 | 13.54 | 239.70 | 326.11 | 275.80 | 1.36 |
| Eu154 | 8.59 | 108.83 | 176.53 | 136.31 | 1.62 |
| Kr85 | 10.76 | 51.61 | 76.05 | 61.66 | 1.47 |
| Ra226 | 1420 | 264.65 | 266.12 | 265.35 | 1.01 |
| Sr90 | 29.04 | 71.27 | 82.19 | 75.89 | 1.15 |

We remind the reader that the current resulting from the decays in the Ra226 source derives from all of the daughters of that nuclide, including some beta-decaying daughters, nearly all of which are in secular equilibrium with the parent.

## 3. DATA PLOTS AFTER PROCESSING

For each nuclide, we find the best-fit exponential and then use that fit to detrend and normalize the measurements, as follows. From the times $t_n$ and current measurements $x_n$, we form

$$V = \sum_n \left( \log\left( x_n \right) + \kappa\, t_n - C \right)^2 \tag{1}$$

and determine the values of $\kappa$ and $C$ that minimize $V$. The detrended and normalized values are then given by

$$z_n = x_n / y_n \text{ where } y_n = \exp\left(C - \kappa\, t_n\right).$$  (2)

The results are shown in Figures 8 through 14. We see that the resulting plots are remarkably similar, except for the plot for Ba133 (Figure 9), which is significantly more ragged.

Power-spectrum analysis, using a likelihood approach (Sturrock et al. 2005), gives a very strong power at 1 year$^{-1}$ for each nuclide, as shown in Table 2. The amplitude estimates are quite consistent, with mean 0.081% and standard deviation 0.0072%, indicating that the hypothesis of no annual modulation may be rejected at the $11\sigma$ significance level.

Table 2. For each nuclide, this table lists the fitted half life, and, for the annual oscillation, the power, amplitude, and phase of maximum of the normalized current.

| Nuclide | Half Life Fit (y) | Power | Amplitude | Phase |
|---------|-------------------|-------|-----------|-------|
|         |                   |       |           |       |
| Ag108   | 430               | 125   | 0.000846  | 0.0789 |
| Ba133   | 10.74             | 60    | 0.000680  | 0.0808 |
| Eu152   | 13.54             | 127   | 0.000827  | 0.0713 |
| Eu154   | 8.59              | 122   | 0.000813  | 0.0684 |
| Kr85    | 10.76             | 113   | 0.000741  | 0.0616 |
| Ra226   | 1420              | 122   | 0.000852  | 0.0707 |
| Sr90    | 29.04             | 115   | 0.000877  | 0.0832 |
|         |                   |       |           |       |
|         |                   | mean  | 0.000808  | 0.0735 |
|         |                   | stdev | 0.000072  | 0.0077 |

Estimates of the phase of maximum are also quite consistent, with mean 0.074 and standard deviation 0.008, corresponding to a maximum at January 27, with a standard deviation of only 3 days. These estimates of the phase are consistent with the range expected of a solar origin of an agent influencing beta decays (Sturrock et al. 2011a).

Table 3. For each nuclide, this table lists the fitted half life, and, for the annual oscillation, the power, amplitude, and phase of maximum of the current offset.

| Nuclide | Half Life Fit (y) | Power | Amplitude | Phase |
|---------|-------------------|-------|-----------|-------|
|         |                   |       |           |       |
| Ag108   | 430               | 125   | 0.015     | 0.0789 |
| Ba133   | 10.74             | 58    | 0.0972    | 0.0821 |
| Eu152   | 13.54             | 128   | 0.2282    | 0.0726 |
| Eu154   | 8.59              | 123   | 0.111     | 0.0694 |
| Kr85    | 10.76             | 116   | 0.0448    | 0.0629 |
| Ra226   | 1420              | 122   | 0.226     | 0.0707 |
| Sr90    | 29.04             | 114   | 0.067     | 0.0838 |
|         |                   |       |           |       |
|         |                   | mean  | 0.113     | 0.0743 |
|         |                   | stdev | 0.084     | 0.0075 |

If the annual oscillation were due to an experimental or environmental effect, it should show up in the current measurements. We can therefore check this possibility by carrying out a similar power spectrum analysis of the current measurements. The results are shown in Table 3. We see that the mean amplitude is 0.113, and the standard deviation is 0.084, corresponding to $1.3\sigma$, which is not significant. This result does not support the possibility of an experimental or environmental effect.

## 4 . DISCUSSION

It is notable that the oscillation amplitudes of the detrended and normalized data derived from PTB-1 vary little with time or from nuclide to nuclide, whereas the same is not true of data derived from PTB-2 (Schrader 2010; Sturrock et al. 2014b). According to Dr Schrader, the only difference between the two experiments is that current measurements were made by means of a Townsend-balance for PTB-1, and by means of a Keithley electrometer for PTB-2.

Concerning PTB-1, Dr Schrader (2012) has expressed the opinion that *the anomalies can best be explained either by an interaction of the solar wind or something similar with the decaying radionuclide (changing the decay rate) or by an interaction of solar (or cosmic) particles or other annually changing parameters with the measuring system (including electronics).* Schrader has specifically suggested that the electronics may be influenced by radon gas. However, such an effect would be manifested in the current measurements but, as we have seen from Table 3, the annual modulations of the current measurements are not consistent.

However, great caution is appropriate in analyzing data from such delicate experiments. As an example, we point to the recent article by Unterweger & Fitzgerald (2014) in which the authors report their realization that earlier half-life measurements had been perturbed by the motion of a loose source holder.

Even if we attribute the oscillations to a solar influence on the decay rates, the results are still surprising. The BNL experiment (Alburger et al. 1986) measured the decays of Cl36 and Si32 in the same equipment, yet the amplitudes of the annual oscillations were found to differ (0.057% for Cl36 and 0.027% for Si32; Sturrock et al. 2011a). Furthermore, some decay-rate experiments seem to give no indication of an annual oscillation. (See, for instance, Bellotti et al. (2012, 2013).) These considerations raise the possibility that most measurements of beta-decay rates may have been subject to complicating or disturbing factors that have not played a role in the PTB experiments.

It may therefore be significant that the PTB experiments use a $4\pi$ detector. To the best of our knowledge, this is not true of any other beta-decay experiment. This difference could be significant if a stimulated-beta-decay process is anisotropic – if, for instance, the direction of the emergent electron or gamma photon is related to the direction of the incoming neutrino. If, as an example, the direction of an emergent secondary particle is parallel or close to parallel to the direction of the incoming neutrino, a typical (non-$4\pi$) experiment would detect a stimulated decay only if the direction from the detector back to the source is sufficiently close to the solar direction (which is, of course, a function of both time of day and time of year). We have in fact found evidence of such an anisotropy in our analysis of radon decays (Sturrock et al. 2012a),

The results of the present analysis appear to be compatible with the hypothesis that the decays are influenced by solar neutrinos, as suggested by Falkenberg (2001), Jenkins et al. (2009a,b) and Fischbach et al. (2009). This could explain why the patterns of variation of the decay rates are so stable and so similar, and bear no apparent relationship to the actual values of the current measurements. Furthermore, the ranges of the phases are compatible with our estimates of the ranges to be expected of a solar influence (0 to 0.183 and 0.683 to 1; Sturrock et al. 2011a).

An annual oscillation, similar to that which is evident in Figures 8 through 14, appears also in the Ra226 data acquired by the PTB-2 experiment (Schrader 2010). Jenkins et al. (2009a) analyzed the Ra226 data from both PTB-1 and PTB-2, together with data (the ratio of measurements of Si32 decay data and Cl36 decay data) derived from an experiment at the Brookhaven National Laboratory (BNL; Alburger et al. 1986). This comparison showed that both experiments yield evidence of an annual oscillation in the decay rates. Jenkins et al. (2010) subsequently investigated the possibility that the annual oscillations may have been caused by various possible environmental effects, and found that to be unlikely.

Analyses of beta-decay experiments have also yielded indications of the influence of processes in the solar interior, such as rotation (Javorsek et al. 2011; Parkhomov, 2012; Sturrock et al., 2010, 2012a) and r-mode oscillations (Sturrock et al. 2011b, 2012b, 2013a, 2014a). These effects appear to be compatible with the hypothesis that beta decays may be influenced by solar neutrinos. Within the context of the neutrino scenario, the rotational and r-mode modulations of the beta-decay rates may be attributed to flavor changes caused by the solar internal magnetic field by the RSFP (Resonant Spin Flavor Precession) mechanism (Akhmedov 1988; Lim et al. 1988; Sturrock et al. 2013a,b, 2014a).

These results raise many questions that call for further experimental and theoretical investigations. Experiments that study the possible influence of reactor-produced neutrinos on beta decays would be helpful. Such experiments have been carried out, so far with inconclusive results (de Meijer et al. 2011).

Since processes involved in stimulated beta decay may be anisotropic, the geometrical layout of experiments may well be significant. It would therefore be helpful to have experiments specifically designed for the study of a possible anisotropy, such as a source surrounded by many detectors, or by a detector that moves systematically over a wide range of positions. Similar experiments in the Southern Hemisphere would be interesting. We have pointed out elsewhere that if beta decays are simulated by neutrinos, this process is likely to involve an exchange of energy, which would probably lead to an exchange of momentum. This may lead to a detectable force on a specimen, and possibly to a detectable torque (Sturrock et al. 2013b). There is also a need for long-term experiments designed to determine whether or not there is a relationship between solar activity such as flares and beta-decay anomalies (Jenkins et al. 2009b; Parkhomov 2010b; Bellotti et al. 2013). Another interesting study would be an investigation of the correlation between beta-decay records of two identical experiments as a function of their separation distance. A deep-space experiment, with a significant variation in orbital distance from the Sun, would provide a crucial test of the cause of the annual variation. A key test of the neutrino hypothesis would be the comparison of measurements of beta-decay rates and simultaneous measurements of the solar neutrino flux by a neutrino observatory such as Super-Kamiokande.

It is necessary to bear in mind the possible role of cosmic neutrinos (Jenkins et al. 2009a; Fischbach et al. 2009; Parkhomov 2010a), and also the possibility that there may be a relationship between beta-decay measurements and the dark-matter measurements made by the DAMA/LIBRA Collaboration (Belli et al. 2011; Bernabei et al. 2012a ). It has been suggested that DAMA/LIBRA measurements may be due in part to the beta decay of a small component of K40 (Pradler et al. 2012, 2013; Bernabei et al. 2012b).

The fact that Ba133 measurements have significantly larger scatter than measurements for the other nuclides raises the possibility that it may be subject to some influence other than a solar one. The PTB dataset lists not only the current as a function of time, but also the standard deviation of the current measurements made each day, comprising a representation of the variability of the current. Figure 15 shows a phasegram formed from the Ba133 variability measurements. This shows a feature with phase in the band 0.45 +/- 0.09, corresponding to days 131 to 197 (centered on June 13). Since this is quite different from the phase of 0.081 listed in Table 2, there is reason to suspect that Ba133 measurements reflect some influence in addition to the solar one.

Figure 16 shows, for comparison, the phasegram formed for the annual oscillation of the DAMA/LIBRA data. This shows a feature in the band 0.41 +/- 0.07, corresponding to days 124 to 176 (centered on May 30). As pointed out in the DAMA/LIBRA articles, this phase is compatible with that to be expected of dark matter. This comparison raises intriguing possibilities. Both Ba133 measurements and DAMA/LIBRA measurements may be influenced either by some form of dark matter or by cosmic neutrinos. On the other hand, the two sets of measurements may reflect different but related stimuli.

In addition to further experimental evidence, there is an obvious need for theoretical models of a possible stimulated-beta-decay process. Such a model may require the introduction of a new boson (which, if it exists, may perhaps be termed a "neutrellon"), such as has been proposed, for different reasons, by Stephenson et al. (1996) and Stephenson and Goldman (1993, 1998). It may be helpful to note that analysis of beta-decay data has led to an estimate of $10^{-23.3}$ cm$^2$ for the effective cross section for the collision of solar neutrinos with Si32 nuclei (Sturrock et al. 2014a), suggesting that the cross section for a hypothetical neutrino-neutrino interaction may be far larger than that for neutrino-electron or neutrino-proton interactions, which is of order $10^{-44}$ cm$^2$.

We acknowledge the generous cooperation of Dr Heinrich Schrader of the Physikalisch-Technische Bundesanstalt in providing us with the data analyzed in this article, and with helpful correspondence. We also thank Jeffrey D. Scargle for helpful comments on the analysis.

# REFERENCES

Akhmedov,, E.Kh. 1988, Phys. Lett. B 213, 64.

Alburger, D.E., Harbottle, G., & Norton, E.F. 1986, Earth Planet. Sci. Lett. 78, 168.

Baurov, Yu. A., & Malov, I.F.  2010, Int. J. Pure Appl. Phys. 6, 469.

Baurov, Yu. B., Sobolev, Yu,G., & Ryabov, Yu.V., et al.  2007, Phys. Atom. Nuclei, 70, 1825.

Bellotti, E., Broggini, C., Di Carlo, G., Laubenstein, M., & Menegazzo, R. 2012, Phys. Lett. B 710, 114.

Bellotti, E., Broggini, C., Di Carlo, G., Laubenstein, M., & Menegazzo, R. 2013, Phys. Lett. B 720, 116.

Belli, P., Bernabei, R., Bottino, A., Cappella, F., Cerulli, R., Fornengo, N. & Scopel, S.. 2011, Phys. Rev. D 84, 055014.

Bernabei, R., Belli, P., Montecchia, F., Nozzoli, F., Capella, F., et al. 2012a, J. Mod. Phys., Conf. Series 12, 37.

Bernabei, R., Belli, P., Cappella, F., Caracciolo, V., Cerulli, R., et al., 2012b, arXiv:1210.5501.

de Meijer, R.J., Blaauw, W.D., & Smit, F.D. 2011, App. Rad. Isotop. 69, 320.

Emery, G.T. 1972, Ann. Rev. Nucl. Sci. 22, 165.

Falkenberg, E.D. 2001, Apeiron, 8, No. 2, 32.

Fischbach, E., Buncher, J.B., Gruenwald, J.B., Jenkins, J.H., Krause, D.E., Mattes, J.J., & Newport, J.R., 2009, Space Sci. Rev. 145, 285.

Hahn, H.P., Born, H.J., & Kim, J. 1976, Radiochim. Acta 23, 23.

Javorsek, D., Jenkins, & Mattes, J.J. 2011, ApJ, 737, 65.

Jenkins, J.H., Fischbach, E., Buncher, J.B., Gruenwald, J.T., Krause, D.E., & Mattes, J.J. 2009a, Astropart. Phys. 32, 42.

Jenkins, J.H., & Fischbach, E., 2009b, Astropart. Phys. 31, 407.

Jenkins, J.H., Fischbach, E., Buncher, J.B., et al., 2009a, Astropart. Phys. 32, 42.

Jenkins, J., Mundy, D.W., & Fischbach, E. 2010, Nucl. Inst. Meth. Phys. Res. A 620, 332.

Lim, C. –S. & Marciano, W.J. 1988, Phys. Rev. D 37, 1368.

Lobashev, V.M., Aseev, V.N.A , & Belesev, A.I . 1999, Phys, Letters B 460, 227.

Namiot, V.A., & Shnoll, S.E. 2006, Phys. Lett. A 359, 249.

Norman E., et al. 2001, Phys. Lett. B 519, 15.

Ohtsuki, T., Yuki, H., Muto, M., Kasagi, J., & Ohno, K. 2004, Phys. Rev. Lett. 93, 112501.

Parkhomov, A.G., 2010a, arXiv:1004.1761.

Parkhomov, A.G., 2010b, arXiv:1006.2295.

Parkhomov, A.G., 2011, J. Mod. Phys. 2, 1310.

Parkhomov, A.G., 2012, arXiv:1012.4174

Pradler, J., Singh, B., & Yavin, I. 2012, arXiv:1210.5501.

Pradler, J., & Yavin, I. 2013, arXiv:1210.7548v2.

Reifenschweiler, O. 1998, Radiat. Phys. Chem. 51, 327.

Schrader, H. 2010, Appl. Radiat. Isot. 68, 1583.

Schrader, H. 2012, private communication.

Shnoll, S.E., et al. 1998, Phys. Usp. 41, 1025.

Shnoll, S.E., et al. 2000, Phys. Usp. 43, 205.

Siegert, H., Schrader, H. & Schoetzig, U. 1998, Appl. Radiat. Isot., 49, 9.

Steinitz, G., Kotlarsky, P., & Piatibatova, O. 2013, Geophys. J. Int. 193, 1110.

Stephenson, G.J., & Goldman, T. 1993, arXiv:9309308.

Stephenson, G.J., & Goldman, T. 1998, Phys. Letters B 440, 89.

Stephenson, G.J., Goldman, T., & McKellar, B.H.J. 1996, arXiv:9603392.

Sturrock, P.A., Bertello, L., Fischbach, E., Javorsek II, D., Jenkins, J.H., Kosovichev, A., & Parkhomov, A.G. 2013a, Astropart. Phys. 42, 62.

Sturrock, P.A., Buncher, J.B., Fischbach, E., Gruenwald, J.T., Javorsek, D., Jenkins, Lee, R.H., Mattes, J.J., & Newport, J.R. 2010, Solar Phys. 267, 251.

Sturrock, P.A., Buncher, J.B., Fischbach, E., Javorsek, D., Jenkins, Mattes, J.J. 2011a, ApJ, 737, 65.

Sturrock, P.A., Caldwell, D.O., Scargle, J.D., & Wheatland, M.S. 2005, Phys. Rev. D 72, 113004.

Sturrock, P.A., Fischbach, E., Javorsek II, D., Jenkins, J.H., & Lee, R.H.. 2013b, Proc. 8th Patras Workshop on Axions, WIMPS and WISPS, Chicago and Fermilab: July 18 - 22, 2012.

Sturrock, P.A., Fischbach, E. Javorsek II, D., Jenkins, J.H. Lee, R.H., Nistor, J., & Scargle, J.D., 2014b, Astropart. Phys 59, 47.

Sturrock, P.A., Fischbach, E., & Jenkins, J.H., 2014a, A Possible Role of Neutrinos in Influencing Beta Decays: A Status Summary (in preparation).

Sturrock, P.A., Parkhomov, A., Fischbach, E., 2011b, Solar Phys. 272, 1.

Sturrock, P.A., Parkhomov, A., Fischbach, E., & et al., 2012b, Astropart. Phys. 35, 755.

Sturrock, P.A., Steinitz, G., Fischbach, E., Javorsek, D., & Jenkins, J.H. 2012a, Astropart. Phys. 36, 18.

Unterweger, M.P., & Fitzgerald, R., 2014, App. Rad. Isotop. 87, 92.

## FIGURES

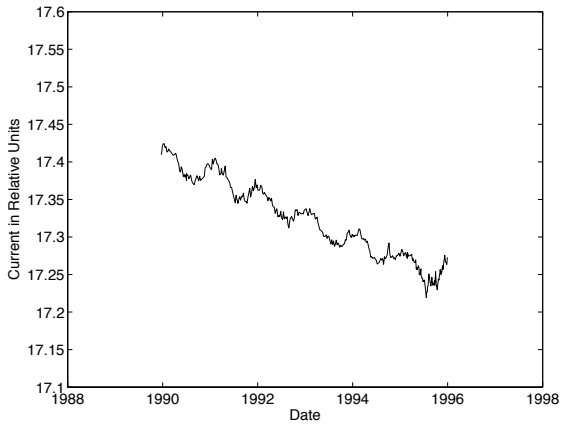

Figure 1. Relative current measurements versus time for Ag108.

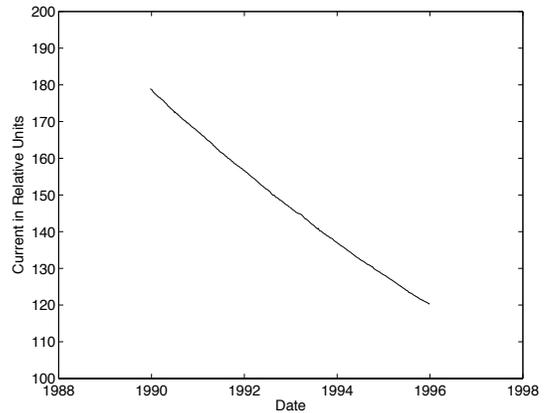

Figure 2. Relative current measurements versus time for Ba133.

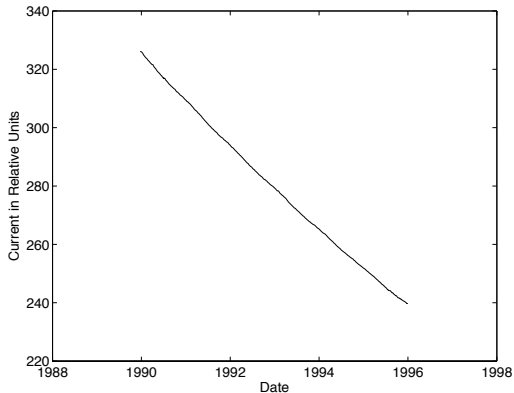

Figure 3. Relative current measurements versus time for Eu152.

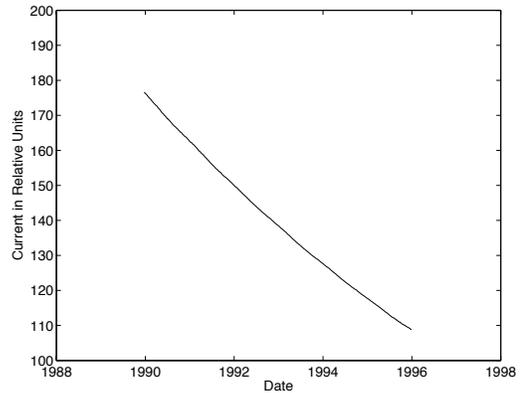

Figure 4. Relative current measurements versus time for Eu154

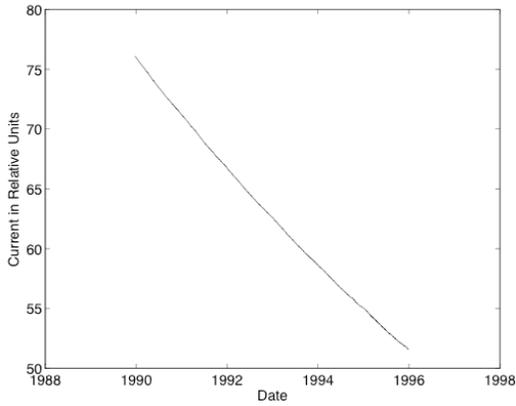

Figure 5. Relative current measurements versus time for Kr85.

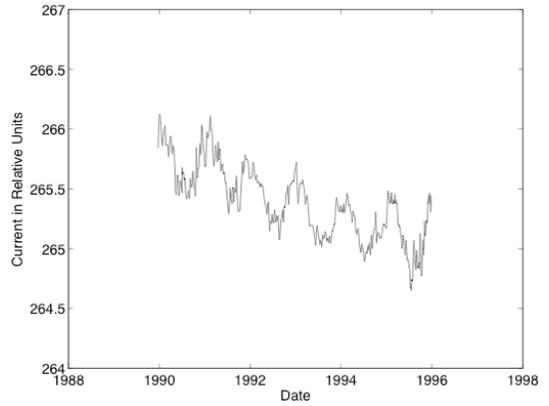

Figure 6. Relative current measurements versus time for Ra226

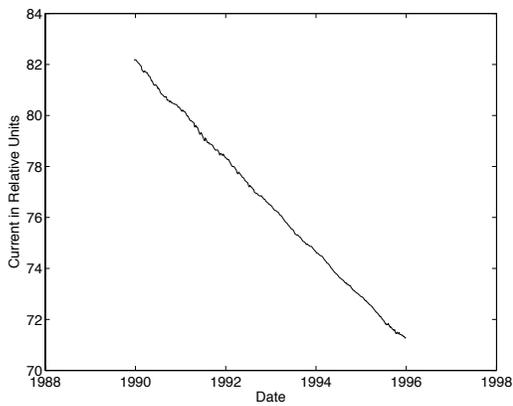

Figure 7. Relative current measurements versus time for Sr90.

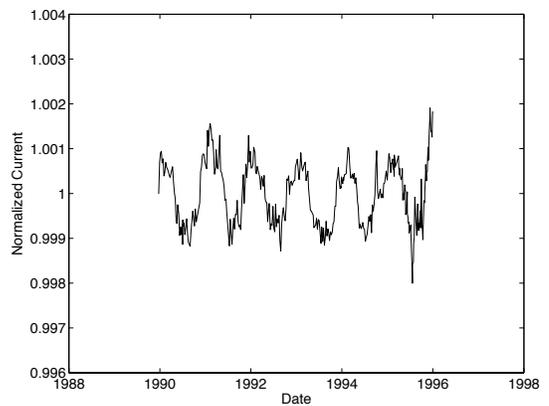

Figure 8. Detrended and normalized relative current measurements versus time for Ag108.

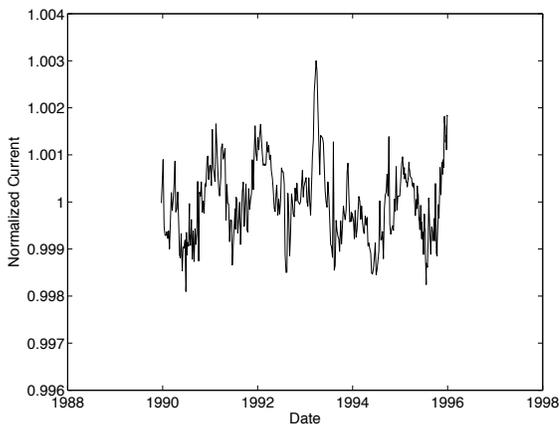

Figure 9. Detrended and normalized relative current measurements versus time for Ba133.

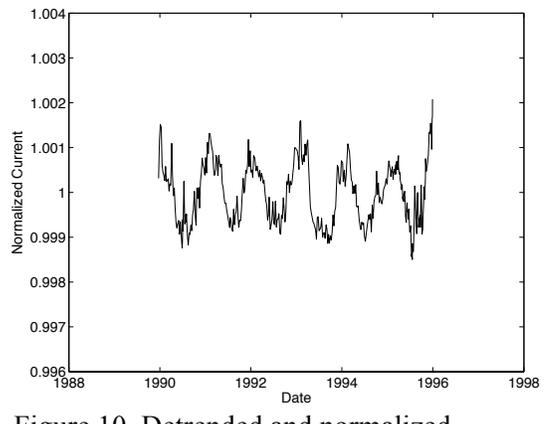

Figure 10. Detrended and normalized relative current measurements versus time for Eu152.

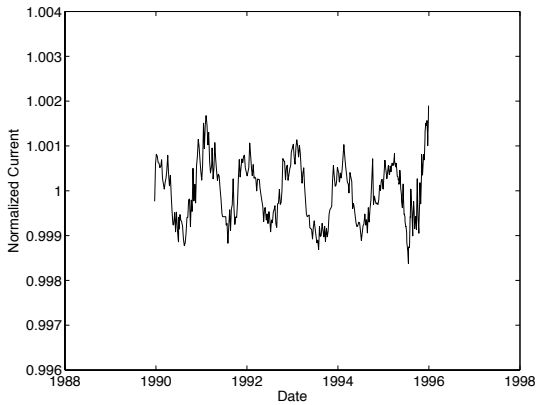

Figure 11. Detrended and normalized relative current measurements versus time for Eu154.

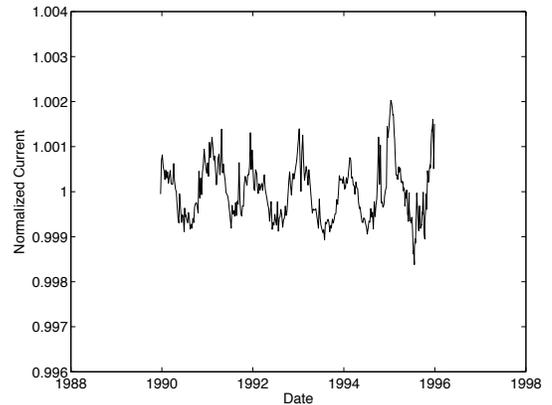

Figure 12. Detrended and normalized relative current measurements versus time for Kr85.

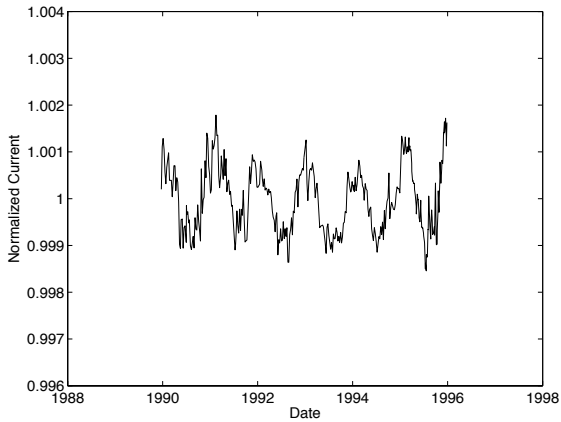

Figure 13. Detrended and normalized relative current measurements versus time for Ra226.

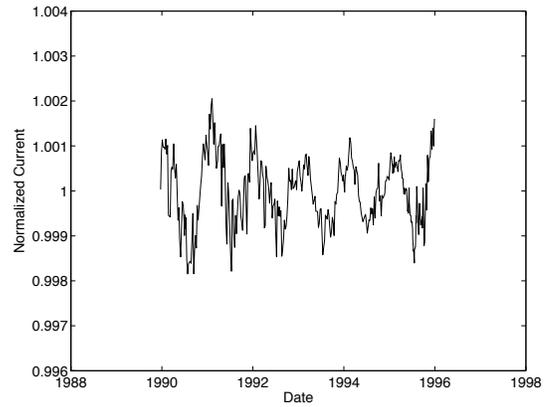

Figure 14. Detrended and normalized relative current measurements versus time for Sr90.

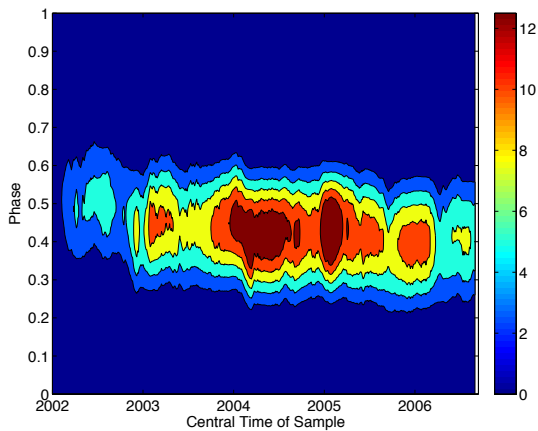

Figure 15. Phasegram formed from variability measures of Ba133 data.

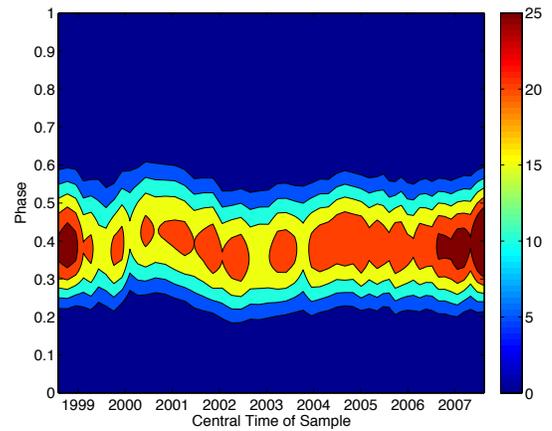

Figure 16. Phasegram formed from DAMA/LIBRA data.